# Symmetry, Entropy, Diversity and (why not?) Quantum Statistics in Society

**J. Rosenblatt**

Institut National de Sciences Appliquées, 20 avenue des Buttes de Coësmes – CS 70839. F35708 Rennes Cedex 7, France; E-Mail: jrosenblatt@wanadoo.fr; Tel.: +33-(0)-1-4655-5574; Fax : +33-(0)-2-2323-8696.

______________________________________________________________________________________
__________

**Abstract.** We describe society as an out-of-equilibrium probabilistic system: $N$ Individuals occupy $W$ *resource* states in it and *produce* entropy $S$ over definite time periods. Resulting thermodynamics is however unusual because a *second* entropy, $H$, measures *inequality* or *diversity* – a typically social feature – in the distribution of available resources. A symmetry phase transition takes place at Gini values $1/3$, where *realistic* distributions become *asymmetric*. Four constraints act on $S$: expectedly, $N$ and $W$, and new ones, *diversity* and *interactions* between individuals; the latter are determined by the coordinates of a single point in the data, the peak. The occupation number of a job is either zero or one, suggesting Fermi-Dirac statistics for employment. Contrariwise, an indefinite number of individuals can occupy a state defined as a quantile of income or of age, so Bose-Einstein statistics may be required. *Indistinguishability* rather than anonymity of individuals and resources is thus needed. *Interactions* between individuals define classes of equivalence that happen to coincide with acceptable definitions of social classes or periods in human life. The entropy $S$ is non-extensive and obtainable from data. Theoretical laws are compared to empirical ones in four different cases of economic or physiological diversity. Acceptable fits are found for all of them.

## 1. Introduction

In previous papers [1], [2], we fitted Lorenz inequality curves [3] – non-thermodynamic quantities at first sight – with a simple model of social entropy. Symmetric distribution laws predict *equal* probabilities of being in the oldest or in the youngest decile, or to belong either to the richest or to the poorest one. In fact, differences between such deciles are found practically everywhere, and this requires, as shown below, Gini [4] coefficients $Gi \geq 1/3$. The assumption of a symmetry phase transition, similar to that in binary alloys [5] and superconductors [6], provides very good fits to data [1]. Four cases in this paper hint indeed at *asymmetric* distributions as the real-world rule.

Entropy can measure social diversity or inequality [7], [8], and perhaps other "qualitative" quantities like difficulty, ability [9] or sensitivity. Here we consider in particular *interactions* between individuals. Inequality indicators should depend on them. Individuals are currently expected to satisfy certain conditions, among them *anonymity* [10], i.e. all permutations of individuals or their resources are equivalent and count for one. We discuss the statistical consequences of this conjecture.

Whether societies are or not in equilibrium is a relevant question in any theoretical approach [11]. They evolve, produce and consume, and therefore we describe them as *nonequilibrium, interacting, entropy producing* and *asymmetrically distributed* statistical systems with a large number of degrees of freedom. We apply the resulting theory to the prediction of data in Ref. 1: they display a rather large variety of fitting parameters and ranges of interaction, necessary to test model forecasts. Two types of data are fitted, economic and demographic, and two examples of diversity are discussed in each of them: incomes in the U.S.A. [12] and *per capita* electricity consumption [13] in 170 countries illustrate the first case, life expectancy up to one hundred years [12] and survival after cancer [14], describe the second one. Data allow a calculation of the average entropy production during periods of five years for cancer, one year in the other cases. Correlations are due to interactions between individuals or similarities in age periods. Dollars, kWh, years of life expectancy or of age without cancer, thus become resource or benefit units (BUs) here. Individuals may refer to persons, households, economic agents, countries, etc. We bridge the gap between individual situations and a global social picture, using the following concepts and their interplay.

(*i*) *States.* The *state* of an *individual* is defined as the amount of resources (of a single type in this paper) available to him or her during a specified period. Our data involve the number of individuals populating a quantile of such states. We could equally well refer to the state of a *benefit*, describing the fraction of total resource allotted to a quantile of the population. Dirac's notation is useful to define individual or benefit configurations: $\langle X|k\rangle$ means "individual $\langle X|$ occupies state $|k\rangle$", while $\langle x|j\rangle$ means "resource $\langle x|$ is allotted to state $|j\rangle$". The number of states does *not* necessarily coincide with that of individuals, $N$ (think of jobs as states and workers as individuals), and the same can be said of the total number of benefit states and total resource $W$. We form groups of states following arbitrary criteria (for example, deciles), or such groups may be *spontaneous*, i.e. socially generated, like the middle class or the adult population. Let $N_k$ be the number of individuals in group $k$ ($k = 1, 2, \ldots K$) and $G_k$ the number of states in it. The average group occupation numbers are $\nu_k = N_k/G_k$, components of a vector $\boldsymbol{\nu}$. Resources $w_k$, divided by the average $\bar{w} = W/N$, provide the components of a vector $\boldsymbol{\omega} = \boldsymbol{w}/\bar{w}$. A connection between $\boldsymbol{\nu}$ and $\boldsymbol{\omega}$ finally leads to the distribution law $\nu(\omega)$.

(*ii*) *Entropy* is a sum over states. In isolated systems, if no other constraint than the obvious one that probabilities add up to unity, it reaches its maximum (equilibrium) value when individuals occupy all accessible states with equal *probability*, i.e. when the distribution is uniform, therefore *symmetric*. A very particular case is equal *resources*, i.e. a $\delta$-function distribution law, usually taken as a reference state for Lorenz curves. Social processes are irreversible, societies actually *produce* entropy $S(\boldsymbol{\nu})$ over a given period of time.

(*iii*) *Inequality* or *diversity* is a specifically social parameter. It is measured here by the extropy [15], i.e. the entropy *produced* [16] by an initially out-of-equilibrium system as it evolves towards equilibrium, $\max(H) - H(\boldsymbol{\omega})$. Here $H(\boldsymbol{\omega})$ provides a measurement of inequality and furnishes a constraint on social entropy production $S(\boldsymbol{\nu})$. The normalised version of the extropy, $1 - H(\boldsymbol{\omega})/\max(H)$, is just the redundancy of information theory [17].

(*iv*) *Symmetry.* Children tend to have a longer life expectancy than their parents, and the poor are more numerous than the rich. Populations therefore have *asymmetrical* nonequilibrium distributions, and this is the case of all situations discussed here. Symmetry is a relevant parameter in the present context.

(*v*) *Correlations.* Specific periods in human life, as well as interactions between individuals, are assumed here to establish correlations between them. Interactions are described as being reflexive, symmetrical and transitive, which is just the definition of classes of equivalence. In fact, they will be seen to coincide with acceptable descriptions of social classes or periods like childhood or oldness. Moreover, members of a class cluster naturally, which implies attractive *intraclass* interactions. *Interclass* correlations and non-additive entropies [18], [19] finally furnish a convenient picture of social systems.

(*vi*) *Indistinguishability.* Statistical descriptions of employment and incomes must be drastically different. A Fermi-Dirac (F-D) statistic applies to employment states, just because the number of individuals on a job is either zero or one. Alternatively, if states are specified as quantiles of income, the upper limit to the amount of benefit in any of them is total resource, which pleads for Bose-Einstein (B-E) statistics. Social and economic laws are thus expected to be invariant against *exchange* [20] – rather than permutation – of two *indistinguishable* – rather than anonymous – individuals or resources.

Mathematical functions are assumed to fulfil the conditions of continuity, differentiability, etc., required to perform indicated operations on them. We mark conceptually important conjectures by the letter "C" followed by an ordinal. Indistinguishability means then (C1) that *social phenomena admit a quantum-like statistical description*. Incidentally, other cases exist where classical entities [21], [22] obey quantum statistics.

Section 2 discusses the relation between social states and entropy. Section 3 dwells on fictitious societies of *independent* individuals, and Section 4 examines an inequality- and interaction-dependent model providing rather good fits to actual data. Conclusions appear in Section 5.

## 2. Symmetry, entropy and universality.

Let $F(\omega)$ be the cumulative population fraction (CPF) and $L(F)$ the cumulative benefit fraction (CBF). The Gini coefficient is, by definition,

$$Gi = 2\int_0^1 (F - L(F))\, dF = 1 - 2\int_0^1 L(F)\, dF = 1 - 2\langle L\rangle. \tag{1}$$

That is, all Lorenz curves having the same value of the constant $\langle L\rangle$ have the same Gini coefficient. Consider now *symmetric* distributions and their Gini-equivalent *uniform* distributions, with maximum and minimum benefits $\omega_M$ and $\omega_m$, respectively. Define $R_u = \omega_m/\omega_M \geq 0$: perfect equality requires $R_u = 1$, maximum inequality has $R_u = 0$. The uniform probability density function is $f_u(\omega) = 1/(\omega_M - \omega_m)$ when $\omega_M \geq \omega \geq \omega_m$, zero otherwise, with CPF $F_u(\omega) = (\omega - \omega_m)/(\omega_M - \omega_m)$. The CBF is $L_u(F_u, R_u) = [2R_u F_u + (1 - R_u)F_u^2]/(1 + R_u)$, implying

$$Gi = \frac{1}{3}\frac{1 - R_u}{1 + R_u} \leq \frac{1}{3}. \tag{2}$$

Symmetric distributions are not only highly improbable, they also have a *maximum* Gini value $1/3$; it is shown in [1] that they *impose* $\omega_m + \omega_M = 2$, while $\omega_M \gg 2$ is quite common in real distributions. Experimental evidence supporting Eq. (2) results from size distributions of beer bubbles [23]. Figure 3 in this reference shows a very great number of Gini coefficients *above* 0.33, and *none* below. Now, since asymmetric distributions do exist, a symmetry change – a phase transition – must take place. In such a case *universality* is expected, whereby near the transition thermodynamic quantities and their possible social counterparts are *generalised homogeneous functions* [24] of their arguments. We apply this condition to social welfare [10].

### 2.1. Welfare, inequality and symmetry.

Social welfare $U(\boldsymbol{w}; W, N)$ and $\boldsymbol{w}$ must increase with $W$ and decrease as $1/N$ when the population increases but total benefit is constant. Generalised homogeneity then means that transformations $W \to aW$ and $N \to bN$ reduce to multiplication of both $U(\cdot)$ and $\boldsymbol{w}$ by $a/b$. With $a = 1/W$ and $b = 1/N$, one finds:

$$U(\pmb{w};W,N) = \frac{b}{a} U\left(\frac{a}{b}\pmb{w}; aW, bN\right) = \frac{W}{N} U\left(\frac{\pmb{w}}{\bar{w}};1,1\right) = \bar{w}\, U_0(\pmb{\omega}), \tag{3}$$

i.e., $U(\cdot)$ is a product of two factors: $\bar{w}$ and $U_0(\pmb{\omega})$, where the independent variable in the latter is necessarily $\pmb{\omega} = \pmb{w}/\bar{w}$. Social welfare should decrease as inequality increases, a condition satisfied by Foster and Sen's [25] proposal, $U(\cdot) = \bar{w}(1-I)$, where $0 \leq I \leq 1$ is a suitable inequality indicator. Their expression coincides with Eq. (3) if $U_0(\pmb{\omega}) = 1 - I(\pmb{\omega}) = H(\pmb{\omega})/\max(H(\pmb{\omega}))$ is the normalised measure of equality. Properties of inequality indicators [10] easily follow from the fact that $\pmb{\omega}$, and therefore $I(\pmb{\omega})$, are scale-, replication- and permutation-invariant, i.e. they do not change if all benefits are multiplied by the same positive constant, the distribution is replaced by a number of replicas of itself, or the ordering of components of the vector $\pmb{\omega}$ is changed. Economical and thermodynamic approaches coincide.

### 2.1.1. Interactions.

A phase transition reveals *interactions* in a thermodynamic system. Assume then that individuals occupy sites $\pmb{r}_i$ in a periodic lattice embedded in a Euclidean space of dimensionality $d$. Interaction links between them are randomly established. We measure distances $r_{ij} = |\pmb{r}_i - \pmb{r}_j|$ in this space in units of nearest-neighbour distance and assume correlations to exist and to decrease as $\mathrm{corr}(\pmb{r}_i, \pmb{r}_j) \sim r_{ij}^{-\delta}$, with $\delta$ positive. Such is the case of percolative clusters. In a qualitative approach [19], consider constant-density groups: an individual in a cluster of linear size $R \sim N^{1/d}$ interacts with $\int_1^R r^{d-1-\delta} dr = (1 - N^{-\theta})/\theta d = \ln_\theta N / d$ other individuals, where $\theta = (\delta/d) - 1$ and $\ln_\theta N \xrightarrow[\theta \to 0]{} \ln N$. We refer to functions $\ln_\theta(\cdot)$ as quasi-logarithms. If $\theta$ is positive, when $N$ goes to infinity the number of interactions per individual is finite, of the order of $1/\theta d$. This defines *short-range* correlations: society behaves as an assembly of noninteracting *finite* clusters. *Long-range* correlations, where each individual is connected to infinitely many others when $N$ grows without limit, occur for $\theta \leq 0$. The parameter $\theta$ thus conveys information on the existence, the range and, as we shall show in Subsection 4.1, the strength of many-body interactions. We point out that $\ln_q(\cdot)$, where $q = \theta + 1 = \delta/d$, is a more usual notation for quasi-logarithms.

## 2.2. Classical independent individuals.

Let us describe society as composed of *noninteracting* and *anonymous* individuals, whose permutations count for one. They form $K$ groups; the entropy is that of Maxwell, Boltzmann and Gibbs (MBG), $S_{MBG} = \ln \Gamma_{MBG} = \ln\left(\sum_{\{N\}} \frac{N!}{N_1! N_2! \ldots N_K!}\right)$. The symbol $\{N\}$ means that *each* term in the sum satisfies $\sum_k N_k = N$. Inequality should be relevant in social systems. Its measure is given by the MBG entropy $H_{MBG} = \ln \Omega_{MBG} = \ln\left(\sum_{\{W\}} \frac{W!}{W_1! W_2! \ldots W_K!}\right)$ in $U_0(\pmb{\omega}) = H(\pmb{\omega})/\max(H(\pmb{\omega}))$, Eq. (3). A simple textbook exercise [20] shows that this is *not* the right way to count configurations in social systems: quantum statistics are necessary, as we now show.

### 2.2.1. Paradoxical distinguishability.

Let an elementary society consist of two *distinguishable* individuals, $A$ and $B$, two equally distinguishable BUs labelled $a$ and $b$, and $G = C = 3$ states, numbered $k = 1, 2, 3$. Employment states result from three jobs that individuals can occupy or not, and where available resources can alight. If states are instead defined by income, the amount of resource in each of them is arbitrary. We use Dirac's notation as discussed in the Introduction. An equal sign relates *equivalent* configurations (all permutations count for one), while the sign "⇔" indicates their *indistinguishability* (the statistic of independent individuals is either F-D or B-E). The MBG expressions imply that $N =$ $W = 2$ such individuals or BUs populate three states in $\Gamma_{MBG} = \Omega_{MBG} = \sum_{\{2\}} \frac{2!}{N_1! N_2! N_3!} = 9$ ways. A paradoxical result in more than one sense, as we now show.

### 2.2.1.1. Employment paradox.

Anonymity assumes that all permutations of individuals $A$ and $B$ are equivalent and *count for one* [10]. Which is one too many for employment states, because $\Gamma_{MBG}$ involves configurations of the type $\langle A|k\rangle + \langle B|k'\rangle = \langle B|k\rangle + \langle A|k'\rangle$, including those where $k = k'$, that is, where individuals $A$ and $B$ occupy the *same* job. The notion of *state* shows here its relevance: anonymity ignores the fundamental zero-or-one restriction on the occupation of employment states. Only three states instead of nine are possible if $\langle A|k\rangle + \langle B|k'\rangle \Leftrightarrow \langle A|k'\rangle + \langle B|k\rangle$ must satisfy the condition $k \neq k'$. This is similar to Gibbs paradox in classical statistical physics. A F-D statistic furnishes the right value for employment; $G_k$ possible states result in $\Gamma_{FD_k} = G_k!/N_k!(G_k - N_k)!$ instead of $\Gamma_{MBG}$.

### 2.2.1.2. Resource paradox.

Five ten-unit banknotes are *physically* distinguishable from a single bill of fifty units, but they are *socially* indistinguishable. Individual states have total benefit as an upper limit of income, so this type of resource obeys B-E statistics. Three states $|k\rangle$ involve therefore six possible configurations instead of nine, of the type $\langle a|k\rangle + \langle b|k'\rangle \Leftrightarrow \langle a|k'\rangle + \langle b|k\rangle \Leftrightarrow \langle 2a|k\,k'\rangle$, where now $k = k'$ is included. Combinatorics gives the number of configurations for $C_k$ benefit states as

$$\Omega_{BE_k} = \frac{(W_k + C_k - 1)!}{W_k!(C_k - 1)!} \approx \frac{(W_k + C_k)!}{W_k! C_k!}. \tag{4}$$

The second Eq. (4) applies when $C_k \gg 1$.

### 2.2.1.3. Individuals' paradox.

Social individuals, like resources, are B-E indistinguishable, and an equation similar to (4) should apply. Indeed, one finds $\Gamma_{BE} = 6$ for our elementary society. With $G_k$ states and $N_k$ individuals in group $k$, we have:

$$\Gamma_{BE_k} = \frac{(N_k + G_k - 1)!}{N_k!(G_k - 1)!} \approx \frac{(N_k + G_k)!}{N_k! G_k!}. \tag{5}$$

The second equation applies when $G_k \gg 1$. Statistics for elementary particles result from their spin and are therefore an *intrinsic* particle property, but they depend on the *nature* of individual states in social systems. The same individuals may obey F-D employment statistics and display B-E behaviour when their incomes are at stake.

## 2.3. Unattainable dilution.

Is there a connection between the number of states $C_k$ (Eq. (4)) and the number of individuals $N_k$ (Eq. (5)) in spontaneous groups? Classical statistics would require a high degree of dilution, i.e. many more benefit states than individuals, $N_k/C_k \ll 1$. This would mean, for examples discussed here, many more jobs than employees, or life expectancy at birth well above one hundred years. In actual fact, these quantities are not strictly equal but of the same order of magnitude, $C_k \approx N_k = G_k \nu_k$, which asserts the *impossibility* of dilution. We therefore conjecture that the attractive intraclass interaction, referred to in the Introduction – a many-body effect – is strong enough to induce high occupancy of available benefit states. This amounts to a new formulation, C1', of C1 in the particular case of *spontaneous* groups: *Dilution is impossible for such groups. They never behave classically.*

## 3. Entropic duality.

Apply C1' and Stirling's large number formula to Eq. (4). One obtains a dimensionless measure of social diversity using the out-of-equilibrium B-E entropy [26] with $G_k \nu_k$ states in group $k$:

$$H_{BE}(\boldsymbol{\omega}) = \sum_{k=1}^{K} \ln \Omega_{BE_k} = \sum_{k=1}^{K} C_k h_{BE}(\omega_k) \approx \sum_{k=1}^{K} G_k \nu_k h_{BE}(\omega_k), \tag{6}$$

to which the group contribution is

$$h_{BE}(\omega_k) = (1 + \omega_k)\ln(1 + \omega_k) - \omega_k \ln \omega_k. \tag{7}$$

Social individuals produce in turn $\boldsymbol{\nu}$-dependent entropy resulting from Eq. (5):

$$S_{BE}(\boldsymbol{\nu}) = \sum_{k=1}^{K} \ln \Gamma_{BE_k} = \sum_{k=1}^{K} G_k s_{BE}(\nu_k), \tag{8}$$

where the B-E entropy production by $\nu_k$ individuals on $G_k$ states is

$$s_{BE}(\nu_k) = (1 + \nu_k)\ln(1 + \nu_k) - \nu_k \ln \nu_k. \tag{9}$$

Equation (9) applies equally well to equilibrium and nonequilibrium entropies, but of course the values of $\nu_k$ in each case are different. The functional relation $\nu(\omega)$ requires still another conjecture, C2: *The most probable path for entropy production maximises the number of ways of reaching the final distribution, and thereby the socially constrained entropy production $S_{BE}(\boldsymbol{\nu})$ during the period of interest* [27].

### 3.1. Constraints.

Two constraints are obvious, $N = \sum_{k=1}^{K} G_k \nu_k$ and $W/\overline{w} = \sum_{k=1}^{K} G_k \nu_k \omega_k$. Lagrange multipliers should thereby result in two adjustable parameters, namely $\alpha$ and $\beta$. But the entropic form $H_{BE}(\boldsymbol{\omega})$, due to society's self-inflicted inequality, is a *third* constraint. An additional Lagrange multiplier is necessary, and results in a new parameter, $\lambda$, which measures diversity. According to C2 and Eq. (6) to (9), entropy production $s_{BE}(\nu_k)$ obeys

$$\frac{\partial}{\partial \nu_k}[s_{BE}(\nu_k) - \nu_k(\beta \varphi_{BE}(\omega_k, \lambda) + \alpha)] = \ln\left(1 + \frac{1}{\nu_k}\right) - [\beta \varphi_{BE}(\omega_k, \lambda) + \alpha] = 0, \tag{10}$$

i.e. the first term in the second Eq. (10) is a linear function of the social free energy per individual:

$$\varphi_{BE}(\omega, \lambda) = \omega + \lambda[(1 + \omega)\ln(1 + \omega) - \omega \ln \omega], \tag{11}$$

formally similar to the Helmholtz free energy per molecule of a B-E ideal gas at "temperature" $-\lambda$; as shown in the next Subsection, this quantity is positive. The distribution law for *noninteracting* individuals is:

$$\nu_{BE}(\omega; \alpha, \beta, \lambda) = \frac{1}{\exp(\beta \varphi_{BE}(\omega, \lambda) + \alpha) - 1}. \tag{12}$$

In case of a F-D statistic for individuals, but no change in the B-E nature of resources, it becomes

$$\nu_{FD}(\omega; \alpha, \beta, \lambda) = \frac{1}{\exp(\beta \varphi_{BE}(\omega, \lambda) + \alpha) + 1}. \tag{13}$$

A similar procedure would apply to any number of independent resources, as many entropic forms $H_{BE,i}$, parameters $\lambda_i$ and several peaks in the distribution. In the particular case of our data, they show a *single* peak in $\nu_{BE}(\omega)$ at $\omega = \omega_p$: it is a *poverty peak* for income or electrical consumption, a *youth* peak for life expectancy and an *old-age* peak for cancer incidence. It coincides with a minimum of the social free energy.

### 3.2. Parameters

Since $\varphi_{BE}(0, \lambda) = 0$ for any finite $\lambda$, $\alpha > 0$ determines the fraction of population $1/(e^\alpha - 1)$ suffering from *extreme* poverty or very short life expectancy, i.e. $\omega \approx 0$. Now, $\omega$ must be positive because nobody can survive without resources; $\alpha = -\beta \mu$, where $\mu < 0$ is the counterpart of the chemical potential in physics. The peak abscissa $\omega_p$ defines $\lambda$ through

$$\omega_p(\lambda) = \frac{1}{e^{-1/\lambda} - 1}, \quad \lambda(\omega_p) = -\frac{1}{\left.\frac{dh_{BE}}{d\omega}\right|_{\omega=\omega_p}} = -\frac{1}{\ln\left(1 + \frac{1}{\omega_p}\right)}. \tag{14}$$

Since $\omega_p$ is positive, only negative values of $\lambda(\omega_p)$ are realistic. Parameters $\beta$ and $\alpha$ result from linear regression once $\lambda$ has been determined from Eq. (14); $1/\beta = \langle\omega\rangle + \lambda\langle h_{BE}(\omega)\rangle = \langle\varphi_{BE}(\omega, \lambda)\rangle$ plays the role of absolute temperature.

### 3.3. Anonymity.

In the classical limit, the additional constraint becomes $H_{MBG}(\omega_k) = \sum_k^K G_k \nu_k \omega_k (1 - \ln \omega_k)$. In place of Eq. (12) one obtains $\nu_{MBG}(\omega, \lambda_{MBG}) = \exp\{-[\beta_{MBG} \varphi_{MBG}(\omega, \lambda_{MBG}) + \alpha_{MBG}]\}$. Entropy production is given in such a case by $S_{MBG}(\nu_k) = \sum_k^K G_k \nu_k (1 - \ln \nu_k)$. We get, in place of Eq. (9), $\varphi_{MBG}(\omega, \lambda_{MBG}) = \omega(1 + \lambda_{MBG} \ln \omega)$. The poverty peak is found at $\omega_{MBG,p} = \exp(1/\lambda_{MBG})$.

## 4. Results.

The distribution of incomes in U.S.A. shows apparently spurious oscillations, with local maxima and minima that happen to coincide with tax return brackets. Numerical smoothing was necessary, and it was applied to all distributions to warrant equality of treatment. For the same reason fitting was also sought for smoothed curves. It had anyway little effect on resources other than incomes. Equation (10) assumes *independent* individuals and refers to the *whole* distribution. If they were indeed independent, it should predict a *single* straight line for plots like those in Fig. 1. But (a) displays three such lines, and four segments appear in (b). Not shown, life expectancy also displays four segments, with age boundaries close to those in (b) but in reverse order, 13, 45 and 65 years. Electricity consumption requires five segments, to be discussed in Subsection 4.2.1.

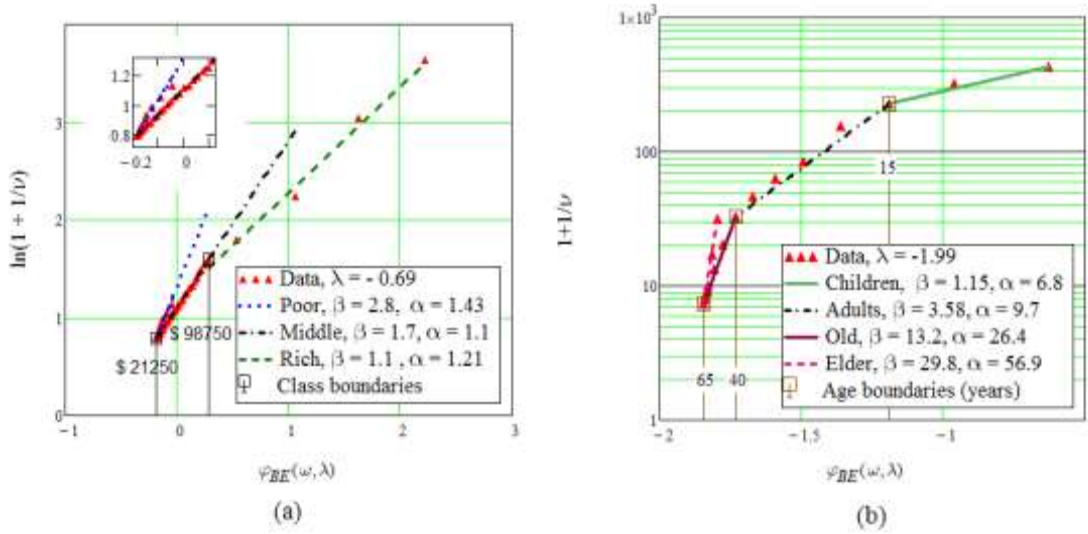

**Figure 1**. Two fits by Eq. (10) of smoothed data on: (a) Incomes in the U.S.A. The plot in the insert is an enlargement of the low-income region. Middle-class boundary incomes are shown. (b) Cancer incidence on male population in New York City, showing relevant ages. The age of occurrence is in fact the benefit (the later, the better) in five-year intervals. Symbol meanings and parameter values appear in the inserts.

In Fig. 1, segments apparently define social classes in (a) and characteristic age periods in (b) through a *piecewise* fit of Eq. (10). Middle-class boundary incomes and relevant periods in human life are indeed credible. Social interactions are not only *short-range*, as described in Subsection 2.1.1., but different classes have different slopes when ordered by increasing benefit. In fact, the definition of absolute temperature that follows Eq. (14), implies that segments fitting "hotter" fractions of society correspond to populations economically or physiologically wealthier, and display less important slopes, as is indeed the case in Fig. 1. Segment slopes are therefore related to *intraclass* interactions. We examine the possibility of *interclass* interactions in the next subsection. The fraction of population $F_p = \Pr(\omega \leq \omega_p)$ *objectively* defines the poorest or the oldest in the distribution. *Dissimilar* data thus reveal *similar* behaviours and plead for a common treatment of different types of diversity.

### 4.1. Interacting classes.

Can the theory feature slope changes? Consider, for instance, a three-class system like that in Fig. 1(a), and originally *independent* quantities $x, y, z$, with probabilities $p_x$, $p_y$, $p_z$, respectively. Total probability $p_{xyz} = p_x p_y p_z$ implies absence of interactions, i.e. $\theta = 0$. Only in this case one would obtain a *single* straight line as predicted by Eq. (10). The substitution of logarithms by quasi-logarithms results in products that should be responsive to interclass correlations:

$$\ln_\theta(p_{xyz}) = \ln_\theta p_x + \ln_\theta p_y + \ln_\theta p_z + (-\theta)[\ln_\theta p_x \ln_\theta p_y + \ln_\theta p_x \ln_\theta p_z + \ln_\theta p_y \ln_\theta p_z] \quad (15)$$
$$+ (-\theta)^2[\ln_\theta p_x \ln_\theta p_y \ln_\theta p_z],$$

where square brackets enclose linear combinations of such products. Factors $(-\theta)^{k-1}$ in Eq. (15) thus measure the strength of a many-body interaction among $k = 1, 2, \ldots, K$ classes.

### 4.2. The model.

An interaction-sensitive model results from the replacement in Eq. (10) of $s_{BE}(\nu_k)$ by $s_\theta(\nu_k) = (1+\nu_k)\ln_\theta(1+\nu_k) - \nu_k \ln_\theta \nu_k$. We have:

$$\frac{\partial}{\partial \nu_k}[s_\theta(\nu_k) - \nu_k(\beta\varphi_{BE}(\omega_k,\lambda) + \alpha)]$$
$$= (1-\theta)[\ln_\theta(1+\nu_k) - \ln_\theta \nu_k] - [\beta\varphi_{BE}(\omega_k,\lambda) + \alpha] \quad (16)$$
$$= \left[\frac{1-\theta}{\theta}\right][\nu_k^{-\theta} - (1+\nu_k)^{-\theta}] - [\beta\varphi_{BE}(\omega_k,\lambda) + \alpha] = 0.$$

Since resources are not expected to interact, Eq. (6) and (14) provide again the measure of equality and the value of $\lambda$, respectively. Plots of $\partial s_\theta(\nu_k)/\partial \nu_k$ as a function of $\varphi_{BE}(\omega,\lambda)$, are close to a *single* straight line. We therefore obtain $\theta$ from the condition that it maximises Pearson's correlation coefficient for this line; $\beta$ and $\alpha$ follow from linear regression on Eq. (16) once $\lambda$ and $\theta$ are determined. Fits of four empirical curves appear in Fig. 2. Class boundaries and periods in human life are obtained from plots like those in Fig. 1. Poverty, for instance, is objectively defined by the region $[0, \omega_p]$ under the data curve in Fig. 2 (a). Fits validate the entropic model as well as conjectures C1, C1' and C2.

#### 4.2.1. Resource-dependent interactions.

Annual *per capita* electricity consumption in Fig. 2(d) is a special case: it is an example of long-range interactions, and it requires *two* values of the $\theta$-parameter, $\theta = -0.18$ for an overall fit and $\theta = 0$ noninteracting behaviour for two groups, one of 22 and the other of 23 countries. A possible explanation is that interactions between countries result mainly from their *exchange* of electricity, often carried out to optimise each country's production systems. The poorest nations rely heavily on – and therefore interact strongly with – other people's production, but correlations disappear as countries become self-sufficient: they form the first group. Increasing production makes trade and therefore correlations to reappear, but they vanish again for the second group, where import and export would compensate each other. The interplay of production, consumption and exchange imposes resource-dependent interactions and thereby several values of the interaction parameter $\theta$.

Figure 2(d) suffers from another drawback: no data exist for the poorest countries (about 20 in number). As a result, the poverty peak is missing. It is assumed here to coincide with the lowest electricity consumption. This suffices to furnish a rather acceptable fit of data.

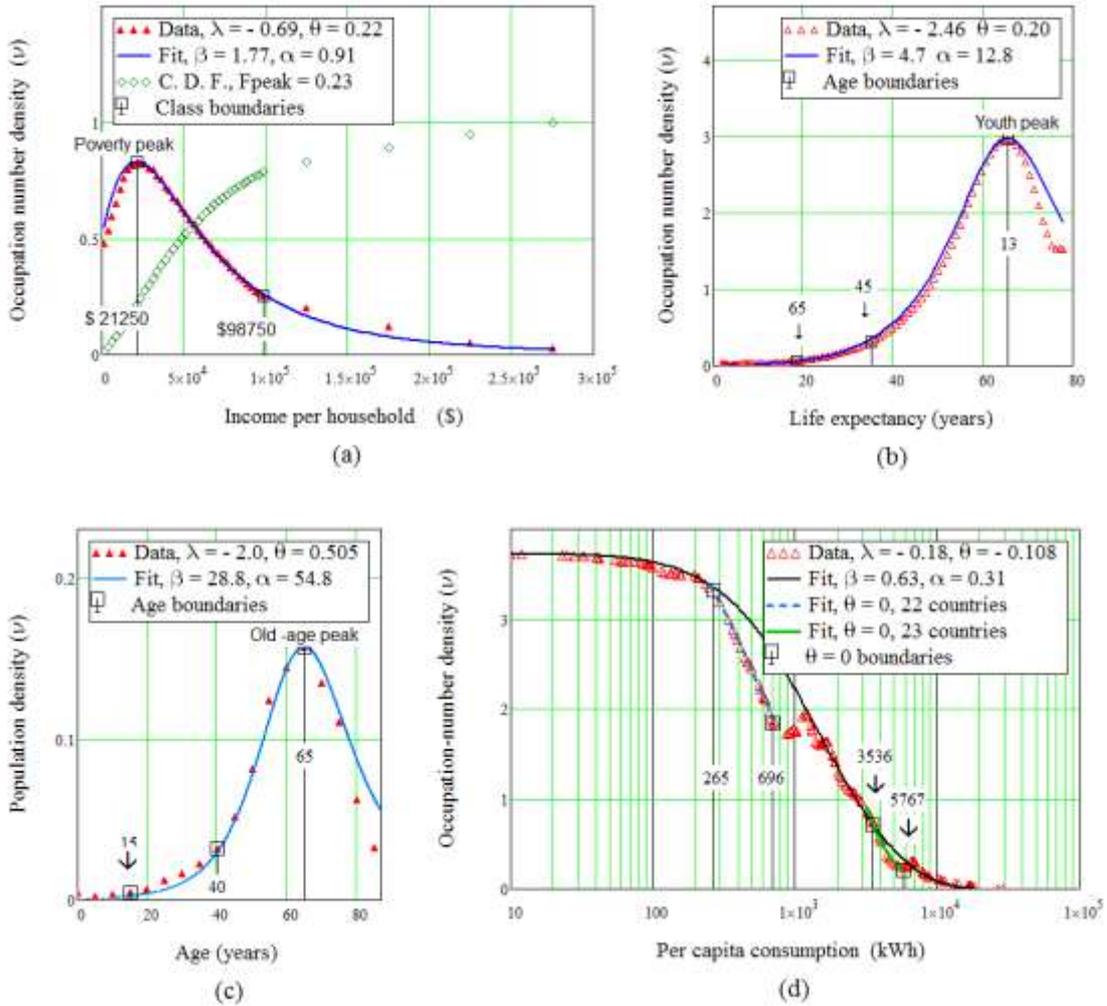

**Figure 2.** Empirical data and theoretical fits for: (a) Income distribution and cumulative distribution function in U. S. A. (b) Life expectancy in U. S. A. (c) New York City cancer incidence. The "benefit" is the age of occurrence, i.e. as late as possible. (d) Electricity consumption per capita in 170 countries. Boundaries result from segmented plots as in Fig. 1. In the first three cases they correspond to class or age limits; in (d), they define intervals where there is no apparent interaction.

## 5. Conclusions.

A single probabilistic model fits data from two types of statistical phenomena, demographic and economic, through four different examples. It depends on the statistics of *state* occupation (of two types, F-D or, as is the case of the four examples in this paper, B-E), the resulting *symmetry* (two possibilities, though symmetric distributions should be extremely unlikely), the type of *interactions* (two levels in this paper, intraclass and interclass), their *intensity* (possibly variable, as for electric energy consumption) and their long or short *range*. Which means about $2^4 = 16$ possible descriptions if symmetry is left aside.

Results on welfare and universality support the idea of a symmetry phase transition at $Gi = 1/3$ between conceivable but unlikely symmetric distributions and realistic asymmetric laws. Equation (16) provides a theoretical expression for distribution laws in societies, whereby specific regions under the distribution curve *objectively* define youth and oldness or poverty and wealth. Different forms of inequality, social and physiological, are thus found to admit the same description.

We obtain good-quality fits to data by applying the paraphernalia of well-known, century-old statistical mechanics (states, entropy, Lagrange multipliers …) to social matters. The *manner* in which this is done is however atypical, particularly because symmetry is a relevant parameter, two entropies at least are at work, and quantum statistics are applied to nonquantum individuals. Societies are considered as nonequilibrium, interacting, entropy-producing statistical systems. As a result, (1) Individuals interact in at least two ways: intraclass and interclass. (2) Inequality, or diversity, is an example of a "qualitative" quantity, for which generalisations of the present approach may be expected. (3) Individuals and resources are clearly not quantal, but *socially* indistinguishable, wherefrom quantum statistics finally follow. This behaviour is not intrinsic, but results from the *nature* of occupied states. (4) One of the entropies is the outcome of an evolving society, the other simply measures inequality in the distribution of available resources, and furnishes a constraint on the former. In the general case, multiple entropies would be required to account for different types of inequality, and as many constraints on social entropy production $S(\nu)$ would result. Conjectures similar to C1' and C2 would be required.

The concepts of extropy, class interaction, multiple entropies and social free energy appear as efficient approaches to nonequilibrium evolving systems. Furthermore, diversity occurs in so many domains that similar methods may be expected to apply to energy production, environmental and other complex systems.

Only two supplementary parameters, $\theta$ due to the strength and range of interactions and $\lambda$ related to inequality, suffice to transform the ideal-gas description of independent individuals into a *predictive* model of society. They result from the coordinates of a single point on the empirical distribution law, the peak. The additional information thus obtained may look rather scanty at first sight, were it not for a remark by E. T. Jaynes [28]: "Entropy as a concept may be regarded as a measure of our degree of ignorance as to the state of a system". Our successful maximisation of entropy production implies then the safest possible assumption, i.e. *minimum* social knowledge of economic and demographic facts.

## Acknowledgements

The author is grateful to Robert Ayres for his relevant questions and to Katalin Martinás for her useful comments.